\documentclass[12pt,preprint]{aastex}

\usepackage{emulateapj5}
\usepackage{apjfonts}

\newcommand{\aboo}{\mbox{$\alpha$~Boo}}
\newcommand{\muHz}{\mbox{$\mu$Hz}}
\newcommand{\muhz}{\mbox{$\mu$Hz}}

\shorttitle{Oscillations in Arcturus}
\shortauthors{Retter et al.}

\begin{document}


\title {Oscillations in Arcturus from {\em WIRE} photometry}





\author{Alon Retter\altaffilmark{1}, Timothy R. Bedding\altaffilmark{1}, Derek L. Buzasi\altaffilmark{2},
Hans Kjeldsen\altaffilmark{3} and L\'aszl\'o L. Kiss\altaffilmark{1}}








\vskip 0.4 cm

\altaffiltext{1}{School of Physics, University of Sydney, 2006, Australia;
retter,bedding,laszlo@physics.usyd.edu.au}
\altaffiltext{2}{Department of Physics, 2354 Fairchild Drive, US Air Force Academy, 
CO 80840, USA; Derek.Buzasi@usafa.af.mil}
\altaffiltext{3}{Theoretical Astrophysics Center, University of Aarhus, 8000 Aarhus C, 
Denmark; hans@ifa.au.dk}









\begin{abstract}

Observations of the red giant Arcturus ($\alpha$ Boo) obtained with the 
star tracker on the Wide Field Infrared Explorer ({\em WIRE}) satellite 
during a baseline of 19 successive days in 2000 July-August are analysed.
The amplitude spectrum has a significant excess of power at low-frequencies. 
The highest peak is at $\sim$ 4.1 $\muhz$ (2.8 d), which is in agreement 
with previous ground-based radial velocity studies. The variability of 
Arcturus can be explained by sound waves, but it is not clear whether these 
are coherent p-mode oscillations or a single mode with a short life-time.



\end{abstract}


\section{Introduction}


Recent results \citep{BBK2001, BC2001, FCA2002} supplied the first strong 
evidence for solar-like oscillations in stars and opened a new era in this 
field. In principle, asteroseismology has the potential to determine the 
basic parameters of the stars such as mass, composition etc. This promise 
motivates extending our knowledge on solar-like oscillations.

$\alpha$ Boo, at V$\sim$0 is the brightest star in the northern hemisphere. 
It is classified as a K1.5 III red giant \citep{KMN1989}. Its effective 
temperature and radius were estimated as $4290\pm30K$ and $\sim 23R_{\odot}$ 
\citep{GLG1999}. Its parallax derived by Hipparcos is 88.85$\pm$0.74 
milli-arc sec \citep{PLK97}, which translates to a distance of 
11.26$\pm$0.09 pc. 


\citet{BJP90a, BJP90b} carried out a radial velocity study of $\aboo$ 
during 82 hours in two weeks in 1988 April-May. They found variability 
at frequencies around several $\muhz$. The highest peak in their power 
spectrum corresponded to 4.3 $\muhz$, with an amplitude of 60 ms$^{-1}$. 
Belmonte et al. listed 19 frequencies and suggested a frequency spacing 
of $\sim$ 5 $\muhz$, although this interpretation was questioned by 
\citet{KB1995}. In addition, Belmonte et al. obtained 18.5 hours of 
photometry of the star during three nights in 1988 April and argued 
that these are consistent with the results of the radial measurements. 
They concluded that the most plausible explanation for the variability 
of $\aboo$ is stellar oscillations.

\citet{HC1994} observed $\aboo$ in radial velocity during eight consecutive 
nights in 1992 June. They found power at frequencies of 4.70, 3.05 and 
1.36 $\muhz$. \citet{M1995} presented three and a half years of intermittent 
radial velocity measurements on Arcturus. These show clearly that it has 
variations of the order of several hundreds ms$^{-1}$ on the long-term. 
While analysis of densely covered parts of the data suggests the presence 
of several periodicities, the strongest at $\sim$ 3 d, there was no 
dominant peak in the power spectrum of the whole data. \citet{M1995} 
concluded that these variations are non-coherent stellar oscillations.


To our knowledge, the only high-precision photometry of Arcturus to date 
is from Hipparcos \citep{Bed2000}. These show that $\aboo$ has optical 
variations with an amplitude of a few percent. Here we present new 
photometric observations of $\aboo$ from the {\em WIRE} spacecraft and 
argue that the variability of Arcturus can indeed be understood by stellar 
oscillations, although it is not clear whether these are coherent p-mode 
oscillations.


\section{Observations and Reduction}

After the failure of the main mission of the Wide Field Infrared Explorer 
({\em WIRE}) satellite, launched by NASA in 1999 March, its star tracker 
was extensively used for asteroseismology of bright objects \citep{BCL2000, 
B2002, PBL2002, CAB2002}. 




\begin{figure*}
\epsscale{0.4} 
\plotone{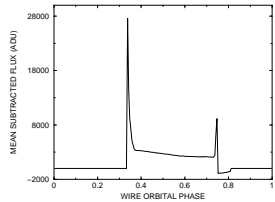}
\caption{The dependence of the instrumental measurements on the orbital 
satellite phase. The spikes at phases 0.34 and 0.75 represent high levels
of scattered light present at the start and end of a typical orbital
segment. Points with $\phi < 0.37$ and $\phi > 0.73$ were thus rejected 
from the analysis and a spline fit to the remaining points was subtracted.}
\end{figure*}

\begin{figure*}
\epsscale{0.4} 
\plotone{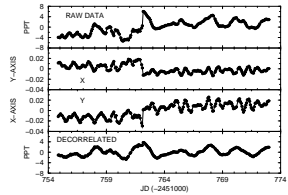}
\caption{Upper panel: The raw light curve of $\aboo$ during the 19-d {\em 
WIRE} run in 2000 July-August. Each point represents a mean of about 4600 
0.1-s exposures obtained during one 96-min orbit.  
Second and third panels: the position of $\aboo$ on the CCD in the X- and 
Y-axis (in pixels) respectively relative to the mean location of the star
in the run. The light curve of $\aboo$ is somewhat affected from its location 
in the CCD. In particular, the fast change at JD=2451762 in the light curve 
of $\aboo$ is most likely instrumental since a similar jump appears in the 
position of the star in the CCD. 
Lower panel: The light curve of $\aboo$ after correcting for the shifts
in the star location. See text for more details.}
\end{figure*}


The observations of $\aboo$ were obtained during 19 successive days in 
2000 July-August using a $512 \times 512$ SITe CCD with 27 $\mu m$ pixels; 
each pixel subtends approximately one arc minute on the sky. The camera is 
unfiltered and the effective response is roughly $V+R$. The observations were 
carried out with a cadence of 0.5 seconds. As described by \citet{Buz2000}, 
data reduction began by applying a simple aperture photometry algorithm, 
which involved summing the central $4 \times 4$ pixel region of the window 
of $8 \times 8$ pixels that can be read. The background level, due 
primarily to scattered light from the bright Earth, was estimated from the 
four corner pixels of each image, and subtracted. We note that $\aboo$  
was bright enough to cause overflows of the onboard analog-to-digital 
converter in the central region of the image. However, we verified 
through examination of count histograms for individual CCD pixels that 
the CCD itself was never saturated, and we were thus easily able to 
reconstruct the actual count rates. A total of 1,284,164 data points 
were acquired during the observing run. 


Following aperture photometry, a clipping algorithm was applied to the 
time series, resulting in the rejection of any points deviating more than 
$2.5\sigma$ from the mean flux, image centroid, or background level. In
practice, at this point in the data reduction process, the rms noise in
the signal was extremely large (about 25 parts per thousand (ppt)) due to 
incomplete subtraction of scattered light, so clipping in this way only 
removes truly deviant points. The majority of these were due either to 
data from the wrong $8 \times 8$ CCD window being returned, or to data 
acquired before the spacecraft pointing had settled during a particular 
orbital segment. 

The crude scattered light removal discussed above is clearly inadequate, 
particularly for observations made early in each orbital segment when 
the target was close to the limb of the bright Earth and the background 
level was therefore especially high. In fact, much of the difficulty 
inherent in the reduction of {\em WIRE} photometric data is in successfully 
characterizing and removing the scattered light signature \citep{Buz2000}. 
In this case, we phased the background data at the satellite 
orbital period (96 min) and binned the result in steps of phase $10^{-3}$. 
The result is shown in Fig.~1, where the zero point for orbital phase is 
arbitrarily chosen. The spikes at $\phi$=0.34 and $\phi$=0.75 represent 
the start and end, respectively, of observations during the orbit. Note 
that points with large scattered light levels are confined to a small 
range of orbital phase ($0.34<\phi<0.37$ and $0.73<\phi<0.81$) and 
that these points represent only a small fraction of the observations, 
especially since the majority of orbital segments do not even contain 
observations made at these phases. Accordingly, we simply discarded 
points with $\phi<0.37$ and $\phi>0.73$. For the remaining points, we 
fitted a spline to the relation shown in Fig.~1 and removed it from the 
data. The resulting time series contained 1,163,420 points, with an rms 
noise level reduced by an order of magnitude, to 2.5 ppt.

\section{Analysis}

The upper panel of Fig.~2 presents the light curve of Arcturus. The data from 
each orbit have been binned into a single mean value, resulting in 252 final 
points. At JD=2451762 there was a jump of about 5.5 ppt in the brightness 
of $\aboo$, however, at this time the location of the centroid of the star 
on the CCD was shifted relatively significantly (Fig.~2, two middle panels). 
The jump is, therefore, instrumental. A further comparison between the light 
curve and the position of $\aboo$ on the CCD suggests that some of the 
wiggles in the light curve are instrumental. In particular, the light curves 
of single satellite orbits typically show linear trends of about 1 ppt, 
which were definitely correlated with the location of the star on the CCD.
These effects are presumably due to sub-pixel structure in the CCD sensitivity.

To remove the effect that minor shifts in the location of the star in the 
CCD have on the light curve, we used the decorrelation method \citep{BGN1991,
RWB1995}. We assumed that the brightness of $\aboo$ was constant during 
single satellite orbits and that the short-term light changes were purely 
an instrumental effect. Using only the data from the first orbit we fitted 
a linearized dependence between the coordinates of the star centroid and 
brightness as $f(x,y)=a+bx+cy$, where $f(x,y)$ is the measured signal, $a$ 
is the assumed constant flux of the star and $b$ and $c$ are the instrumental 
coefficients. By subtracting $bx+cy$ from the data in all other orbits , we 
largely remove the instrumental effect. The result is shown in the bottom 
panel of Fig.~2. It is clear that most of the dependence in the location of 
the star has disappeared and that $\aboo$ shows changes on time scales of 
several days. We note that using different orbits to measure the coefficients 
or adding the higher term ($dx^2+ey^2+fxy$) to the above function had 
negligible effect on the decorrelated light curve. 





The upper panel of Fig.~3 displays the amplitude spectrum of the 
observations after subtracting a linear term from the raw data. The second 
and third panels represent the amplitude spectrum of the shifts in the x- 
and y-axis respectively. The fourth panel in Fig.~3 shows the amplitude 
spectrum of the decorrelated data. There is an excess of power at low 
frequencies, which is centered around 0.35 d$^{-1}$ (4.07 \muhz). This 
periodicity reflects the obvious changes in the light curve (Fig.~2). 




\begin{figure*}
\epsscale{0.4} 
\plotone{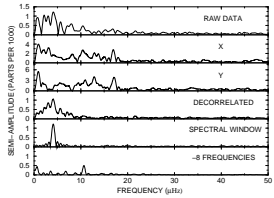}
\caption{
Upper Panel -- Amplitude spectrum of the raw {\em WIRE} data. 
Second and third panels -- The amplitude spectrum of the shifts in the 
CCD of the x- and y-axis respectively.
Fourth panel -- Amplitude spectrum of the corrected data. There is a 
significant excess of power at low frequencies. The four highest peaks 
correspond to the frequencies 4.07, 3.09, 2.44 and 5.71 $\mu$Hz 
respectively. 
Fifth panel -- The spectral window (after planting a 4.07-$\mu$Hz sinusoid 
in the window function with the same amplitude it has in the data). 
Lower panel -- after prewhitening the eight strongest frequencies.}
\end{figure*}


We analysed the amplitude spectrum using the conventional method of 
iterative sine-wave fitting. A synthetic light curve was built using a 
sinusoid at a frequency of 4.07-$\mu$Hz, with the same amplitude as in the 
data and sampled according to the window function (Fig.~3, fifth panel from 
top). We subtracted the 4.07-$\mu$Hz signal from the data and calculated 
the amplitude spectrum of the residuals. The highest peak was at 0.27 
d$^{-1}$ (3.09 $\mu$Hz). We then returned to the original data, fitted 
simultaneously these two frequencies (4.07 and 3.09-$\mu$Hz) and found the 
third highest peak in the residual amplitude spectrum. This process was 
repeated for ten iterations. The result after subtracting eight peaks is 
shown in the bottom panel of Fig.~3. There is still some excess of power 
at low frequencies. Table~1 lists the ten highest peaks in the amplitude 
spectrum found by simultaneous fitting of the frequencies. The typical 
errors on the frequencies are $\sim$ 0.1 $\mu$Hz. Peak 5 (1.56 $\mu$Hz) 
is questionable, since the data extend over only $\sim$ 2.5 cycles.


\begin{table*}
\begin{center}
\caption{The highest peaks in the amplitude spectrum}
\begin{tabular}{crccc}

Peak & Frequency & Semi-amplitude \\
     & (\muhz)   & (ppt)          \\
     &           &                \\
1    & 4.07      & 1.39           \\
2    & 3.09      & 0.98           \\
3    & 2.44      & 0.78           \\
4    & 5.71      & 0.56           \\
5    & 1.56      & 0.47           \\
6    & 4.57      & 0.36           \\
7    & 7.91      & 0.31           \\
8    & 7.35      & 0.26           \\
9    & 10.66     & 0.20           \\
10   & 6.81      & 0.14           \\

\end{tabular}
\end{center}
\end{table*}


Fig.~4 presents the power spectrum (amplitude squared) at the frequency 
range of 0-20 $\mu$Hz. The dotted vertical lines show the locations of 
the ten highest peaks. The long-dashed curve is the Lorentzian fit to the 
power excess at low frequency using the maximum -likelihood method given 
by \citet{TF1992}.


\begin{figure*}
\epsscale{0.4} 
\plotone{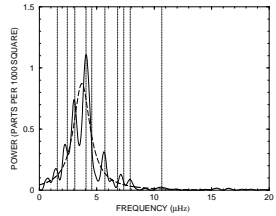}
\caption{Power spectrum of the {\em WIRE} data in the range 0-20 $\mu$Hz. 
The locations of the ten highest peaks are marked by vertical dotted lines. 
The long-dashed curve shows the Lorentzian fit to the envelope of peaks.}
\end{figure*}


\section{Discussion}

The {\em WIRE} photometry clearly shows that $\aboo$ is variable on time 
scales of a few days. There are several frequencies with semi-amplitudes 
of about one ppt or less in the amplitude spectrum. The amplitude and 
frequency of the power excess in Arcturus found in this work are in 
excellent agreement with the predictions for solar-like oscillations 
based on scaling from the Sun \citep{KB1995}. In addition, the shape 
of the power excess is typical of solar-like oscillations \citep{TF1992}
and could be explained by a single mode with a damping life time of about 
2 days (Fig.~4). Alternatively, to establish the presence of p-mode 
oscillations, we must demonstrate a regular series of peaks in the power 
spectrum.

\subsection{Search for a regular series of peaks}

To locate a possible p-mode structure in the $\aboo$ data we first tried to 
fit a simple regular series of peaks to the frequencies we found (assuming 
an asymptotic relation for radial modes). Indeed, the frequencies fit a 
simple regular series having a separation between the modes of 
$\Delta\nu$=0.825$\pm$0.049 \muHz. We note that we assume here that $l=0$ 
is the dominant mode and non-radial modes are negligible, which is expected 
from theoretical models \citep{DGH2001}. Our suggested value for the 
frequency splitting is in excellent agreement with the expected value of 
0.9$\pm$0.2 $\mu$Hz \citep{KB1995} and perhaps consistent with the 
estimate of 0.6 $\mu$Hz reported by \citet{M1995} \footnote{\citet{M1995} 
actually concluded that $\Delta\nu\simeq$~1.2$\,\mu$Hz by assuming that the 
mode l=1 is also present.}, but not with the value of 5.0 $\mu$Hz suggested 
by \citet{BJP90a, BJP90b}. 

Our value of $\Delta\nu$ is, however, close to our frequency resolution 
($\sim$ 0.64 $\mu$Hz), which is set by the length of the run. The same was 
true for the observations by \citet{BJP90a, BJP90b} and \citet{M1995}. 

We carried out simulations to determine whether the regularity in the {\em
WIRE\/} data can be attributed to p-modes.  We generated a simulated time 
series as the sum of two noise sources: white noise (i.e., normally 
distributed) plus non-white noise (a random walk with a high-pass filter).  
We generated five simulations, each with the same sampling as the real data, 
but with different seeds for the random number generator.  We then analyzed 
each noise spectrum in the same way as the real {\em WIRE\/} dataset.

%

All simulations showed a somewhat regular series of peaks, with 6 to 8 peaks 
fitting the model. The value for the separation between the individual peaks 
was only slightly different from the value we found in Arcturus (within a 
half sigma). The scatter on the separation for the simulated time series 
data is also only slightly worse than what is found in Arcturus (within one 
sigma). We therefore conclude that we do not need a regular series of p-mode 
oscillations in order to produce the regularity observed in the {\em WIRE} 
data. We are thus unable, from the present data, to decide whether p-mode 
oscillations exist in Arcturus or not.

The above effect in Arcturus is also able to explain the regular series 
of peaks, with a separation of 2.94 \muhz, observed in $\alpha$ UMa A by
\citet{BCL2000}. A simulation similar to the one above for $\aboo$ indicates 
that one, even in data containing pure noise and with a sampling like the 
one for $\alpha$ UMa A, will find a separation of 3.0$\pm$0.3 $\muhz$ with 
low scatter. This may explain the difference between observations and 
model calculations by \citet{GDB2000}.

\subsection{Granulation noise?}

Could granulation noise, rather than oscillations, be responsible for the
excess power?  This is unlikely, since the variability is also detected 
in velocity, with a semi-amplitude of $\sim$ 60\,m$s^{-1}$ at 4.3\,$\mu$Hz 
\citep{BJP90a, BJP90b}. The ratio of the velocity amplitude to the 
photometric amplitude is about 60 ms$^{-1}$ppt$^{-1}$. This is about twice 
that expected for sound waves \citep[][Eq.~5]{KB1995}, but more than seven 
times greater than observed in solar granulation noise \citep{PRJ1999}.
Also, the power spectrum from granulation noise would be expected to 
continue rising towards low frequencies, which is not observed.

\section{Conclusions}

We strongly established the suggestion that Arcturus is photometrically 
variable. We conclude that the oscillations in Arcturus can be explained 
by sound waves, but it is not clear whether these are coherent p-mode 
oscillations or perhaps a single mode with a short lifetime. The latter 
would explain the different frequencies found by various authors (Section~1). 
A longer time series is required to distinguish between the two options. 


As theory predicts that non-radial modes should be unobservable in red 
giants \citep{DGH2001}, asteroseismology probably will not supply detailed 
information on these objects. Detecting the $l=0$ modes of stellar oscillation 
has, however, the potential of estimating reliably the poorly known masses 
in red giants. Our observations of Arcturus are the first step towards this 
goal.

\acknowledgements

We are grateful to all the people who made the use of the {\em WIRE} 
satellite possible. We thank two anonymous referees for their useful 
comments which significantly improved the paper and Ariel Marom for
fruitful discussions. AR, TRB and LLK are supported by the Australian 
Research Council and HK by the Danish Natural Science Research Council
and the Danish National Research Foundation through its establishment 
of the Theoretical Astrophysics Center. DLB acknowledges support from 
NASA (NAG5-9318).

\end{document}